%
\magnification = \magstep1
\hsize = 16 truecm
\parskip = 0pt
\normalbaselineskip = 17pt plus 0.2pt minus 0.1pt
\baselineskip = \normalbaselineskip
%
%

\font\brm=cmr10 scaled \magstep1
\font\bbf=cmbx10 scaled \magstep1
\font\bit=cmti10 scaled \magstep1



%

\def\Bigbreak{\par \ifdim\lastskip < \bigskipamount \removelastskip \fi
                   \penalty-300 \vskip 10mm plus 5mm minus 2mm}
%
%
\newcount\eqnum
\newcount\tempeq
\def\cleareqnum{\global\eqnum = 0}
\def\eqname{(\the\chnum.\the\eqnum)}
\def\neweq{\global\advance\eqnum by 1 \eqno\eqname}
\def\neweqalign{\global\advance\eqnum by 1 &\eqname}
\def\releq#1{\global\tempeq=\eqnum \advance\tempeq by #1
             (\the\chnum.\the\tempeq)}

\cleareqnum
%
%
\newcount\chnum
\def\clearchnum{\global\chnum = 0}
%
%

\def\newchapt#1{\Bigbreak \global\advance\chnum by 1
                \cleareqnum
                \leftline{\bbf { }#1}
                \nobreak\vskip 5mm plus 2mm minus 1mm}
\clearchnum
%
%
\newcount\notenumber
\def\clearnotenumber{\notenumber=0}
\def\note{\advance\notenumber by1 \footnote{$^{\the\notenumber}$}}
\clearnotenumber
%
%
\newbox\Eqa
\newbox\Eqb
\newbox\Eqc
\newbox\Eqd
\newbox\Eqe
\newbox\Eqf
\newbox\Eqg
\newbox\Eqh
\newbox\Eqi
\def\storeeq#1{\setbox #1=\hbox{\eqname}}

%
%
\pageno=0
\footline={\ifnum\pageno=0\strut\hfil\else\hfil\tenrm\folio\hfil\strut\fi}%
\def\e{{\rm e}}

\def\H{{\cal H}}
\def\D{{\cal D}}

\def\E{{\cal E}}
\def\F{{\cal F}}

\def\P{{\cal P}}

\def\S{{\cal S}}

\def\R{{\bf R}}
\def\C{{\bf C}}

\def\bfR{{\bf R}}
\def\bfC{{\bf C}}

\def\der{\partial }
\def\alg{{\cal A}_R}
\def\balg{{\cal B}_R}

\def\brep{{\cal F}_{R,B}}
\def\hl{{\R_+ } }
\def\at{ {\widetilde a} }
\def\ht{ {\widetilde h} }

\def\supp{ {\rm supp} }
\def\sp{{\rm sp}}

\def\a{a(f)}
\def\ac{a^*(f)}


\rightline {IFUP-TH 47/98}
\centerline {\bbf The Nonlinear Schr\"odinger Equation}
\centerline {\bbf on the Half Line}
\vskip 1 truecm
\centerline {\brm Mario Gattobigio}
\medskip
\centerline {\it Istituto Nazionale di Fisica Nucleare, Sezione di Pisa}
\centerline {\it Dipartimento di Fisica dell'Universit\`a di Pisa,}
\centerline {\it Piazza Torricelli 2, 56100 Pisa, Italy}
\bigskip
\medskip
\centerline {\brm Antonio Liguori}
\medskip
\centerline {\it International School for Advanced Studies,}
\centerline {\it 34014 Trieste, Italy}
\bigskip
\medskip
\centerline {\brm Mihail Mintchev}
\medskip
\centerline {\it Istituto Nazionale di Fisica Nucleare, Sezione di Pisa}
\centerline {\it Dipartimento di Fisica dell'Universit\`a di Pisa,}
\centerline {\it Piazza Torricelli 2, 56100 Pisa, Italy}
\bigskip
\vskip 1 truecm
\centerline {\bit Abstract}
\medskip

The nonlinear Schr\"odinger equation on the half line with 
mixed boundary condition is investigated. After a brief introduction 
to the corresponding classical boundary value problem, 
the exact second quantized solution of the system is constructed. 
The construction is based on a new algebraic structure, which is called 
in what follows boundary algebra and which substitutes, in the presence of 
boundaries, the familiar Zamolodchikov-Faddeev algebra. 
The fundamental quantum field theory properties of 
the solution are established and discussed in detail. 
The relative scattering operator is derived in 
the Haag-Ruelle framework, suitably generalized to the case of 
broken translation invariance in space. 

\bigskip
\centerline {October 1998}
\vfill \eject

\newchapt {I. INTRODUCTION}

The general interest in quantization on the half line
$\hl = \{x\in \R\, :\, x>0\}$ stems from the recently growing
number of applications in different physical areas, including
open string theory, dissipative quantum mechanics and
quantum impurity problems. In the last few years, important
progress has been made in this subject by means of
conformal field theory. Focusing on the nonlinear
Schr\"odinger (NLS) model on $\hl $, in the present paper
we explore the possibility to employ integrability.

Let us recall that when considered on the whole line $\R$,
the NLS model represents one of the most extensively studied
nonrelativistic integrable systems (see e.g. Ref. 1).
The corresponding equation of motion is
$$
(i\der_t + \der_x^2 )\Phi (t,x) = 2g\, |\Phi (t,x)|^2 \Phi (t,x)
\quad ,
\neweq
$$
where $\Phi(t,x)$ is a classical complex field. The model on the half line
is obtained restricting Eq.(1.1) on $\hl $, supplemented
with the boundary condition
$$
\lim_{x \downarrow 0}\left( \der_x - \eta \right)
\Phi (t,x) = 0 \quad . \neweq
$$
Here $\eta $ is a dimensionful parameter of the theory.
{}For $\eta = 0$ and in the limit $\eta \rightarrow \infty $ one
recovers from Eq.(1.2) the familiar Neumann and Dirichlet
boundary conditions respectively.
To our knowledge, the boundary value problem (1.1,2) has
been first investigated by Sklyanin$^2$ and Fokas$^3$, who have shown
that the integrability, which holds for the system on the whole line,
persists also on the half line. Our main goal
below will be to construct the exact second quantized solution
of Eqs.(1.1,2), in the case $g\ge 0$, $\eta \ge 0$.
Concretely, this means:
\medskip 

\item{1.} To construct a Hilbert space $\H_{g,\eta}$ describing the states of
the system;

\item{2.} To define on an appropriate dense domain in $\H_{g,\eta}$
an operator valued distribution $\Phi(t,x)$, $x>0$, satisfying, in a sense that
will be made precise below, the equation of motion (1.1), the boundary
condition (1.2) and the equal time canonical commutation relations
$$
[\Phi(t,x)\, ,\, \Phi(t,y)]= [\Phi^*(t,x)\, ,\, \Phi^*(t,y)]= 0
\quad , \neweq
$$
$$
[\Phi(t,x)\, ,\, \Phi^*(t,y)]= \delta(x-y) \quad , \neweq
$$
where $\Phi^*$ is the Hermitian conjugate of $\Phi $;

\item{3.} To show the existence of a vacuum state $\Omega $ in the 
above mentioned domain, which is cyclic with respect to 
the field $\Phi^*$.
\medskip 

\noindent
The analogous construction in the case of the whole real line has
been carried out some years ago$^{4-9}$ by means of the quantum
inverse scattering transform. The basic algebraic tool of this approach
is the Zamolodchikov-Faddeev$^{10}$ (ZF) algebra $\alg $ - an appropriate
generalization of the canonical commutation relations which incorporates
the two-body scattering matrix $R$. We will show below that
the half line system can be treated in the framework of inverse scattering
as well, the relevant algebraic structure being now
the so called boundary algebra $\balg $. In the same way as the ZF algebra has
been conceived$^{10}$ to represent the factorized scattering of integrable systems
on the line, the general concept of boundary algebra$^{11}$ is inspired by 
Cherednik's scattering theory$^{12}$ of integrable systems on the half line.
The fundamental feature of $\balg $ is that it encodes both the
nontrivial scattering between particles and the reflection from the boundary
at $x=0$.

A preliminary account without proofs, which partially covers the results 
presented below, is given in Ref. 13. This paper is organized as follows. 
In the next section we summarize
some known, but useful facts, about the classical NLS model both on $\R$
and $\hl $. Sec. III represents a summary of those fundamental properties
of $\balg $ and its Fock representations, which are needed in the quantization.
In Sec. IV we define the quantum field $\Phi (t,x)$ and establish its
kinematic properties, verifying the canonical commutation relations (1.3-4).
The dynamics is investigated in Sec. V, where it is shown that Eqs.(1.1,2)
are indeed satisfied. We sketch there also the derivation of the 
correlation functions. Sec. VI is devoted to the asymptotic theory 
of the NLS model on $\hl $. The last section contains our conclusions.

\newchapt {II. THE CLASSICAL NLS MODEL}

The study of the classical NLS equation has a long story.
Without entering the details, we will collect in this section
some basic facts providing useful hints for the quantization.

\bigskip 
\medskip 

\leftline {\bf A. NLS on the real line}

\bigskip

The equation of motion (1.1) on $\R$ is obtained by varying the action
$$
A\left [\Phi , \overline \Phi \right ] =
\int_{\R} dt \int_{\R} dx
\left [ i\overline \Phi (t,x) \der_t \Phi (t,x) -
|\der_x \Phi (t,x)|^2 - g |\Phi (t,x)|^4 \right ]
\quad .  \neweq
$$
The system admits an infinite number of integrals of motion,
the energy
$$
E\left [\Phi , \overline \Phi \right ] = \int_{\R } dx \left [
|\der_x \Phi (t,x)|^2 + g |\Phi (t,x)|^4 \right ]
\neweq
$$
being one of them. Notice
that $E\left [\Phi , \overline \Phi \right ]$
is non-negative as long as $g \geq 0$. This constraint has
an important role in the quantum version of the theory.

About twenty years ago Rosales$^{14}$ discovered that Eq.(1.2)
on $\R$ admits solutions of the form
$$
\Phi(t,x) = \sum_{n=0}^{\infty} (-g)^n \Phi^{(n)}(t,x)
\quad , \neweq
$$
where
$$
\Phi^{(0)}(t,x) \equiv \widetilde \lambda (t,x) = \int_{\R } {dq \over 2\pi}
\lambda (q)\, \e^{ixq - itq^2} \quad , \neweq
$$
solves the free Schr\"odinger equation and
$$
\Phi^{(n)}(t,x) =
$$
$$
\int_{\R^{2n+1}} \prod_{i=1\atop j=0}^n {dp_i \over 2\pi}
{dq_j \over 2\pi}
{\overline \lambda }(p_1)\cdots {\overline \lambda }(p_n)
\lambda (q_n)\cdots \lambda (q_0)
{\e^{i\sum_{j=0}^n (xq_j - tq_j^2 ) - i\sum_{i=1}^n (xp_i - tp_i^2) }
\over \prod_{i=1}^n \left[ (p_i - q_{i-1})\,
 (p_i - q_{i}) \right]} \, . \neweq
$$
The integration in (2.5) is defined by the principal
value prescription and one assumes that $\lambda(k)$ is a function for
which the integrals (2.4,5) exist and the series (2.3) converges uniformly 
in $x$ for sufficiently small $g$. It is not difficult to argue that there is
a large set of such functions; any $\lambda $ belonging to the Schwartz
test function space $\S(\R)$ meets for instance the above requirements.
In fact, expressing $\Phi^{(n)}(t,x)$ in terms of $\widetilde \lambda (t,x)$, one finds
$$
\Phi^{(n)}(t,x) =
\int_{\R^{2n}} \left [\prod_{i=1}^n dy_i dz_i
{\overline {\widetilde \lambda }}(t,y_i) \widetilde \lambda (t,z_i)\right ]
\widetilde \lambda (t,x+\sum_{i=1}^n y_i - z_i)
\sigma (x;y_1,z_1,...,y_n,z_n)
\, , \neweq
$$
where 
$$
\sigma (x;y_1,z_1,...,y_n,z_n) = 4^{-n}\prod_{i=1}^n
\varepsilon \left (x + \sum_{j=1}^{i-1}y_j - \sum_{k=1}^iz_k \right )
\varepsilon \left (\sum_{j=1}^i (y_j - z_j) \right ) \, , \neweq
$$
and $\varepsilon(x)$ denotes the sign of $x$.
Therefore,
$$
|\Phi^{(n)}(t,x)| \leq
\int_{\R^{2n} } \left [\prod_{i=1}^n dy_i dz_i
|\widetilde \lambda (t,y_i)\widetilde \lambda (t,z_i)|\right ]
|\widetilde \lambda (t,x+\sum_{i=1}^n y_i - z_i)| \, . \neweq
$$
At the other hand, using standard estimates one can deduce that for any
$\lambda (k) \in \S(\R )$ there exist two positive constants $\Lambda_1$
and $\Lambda_2$ such that
$$
\int_{\R} dx |\widetilde \lambda (t,x)| \leq \Lambda_1(1+|t|)
\quad , \qquad
\sup_{x\in \R} |\widetilde \lambda (t,x)| \leq \Lambda_2
\quad . \neweq
$$
Combining Eqs.(2.8) and (2.9) we conclude that the series (2.3) 
converges uniformly in $x$ for
$$
g < \left [\Lambda_1(1+|t|)\right ]^{-2} \quad . \neweq
$$

The main reason for focusing on the result of Rosales is because
it turns out$^{5-9}$ that the general structure of the solution (2.3-5) is
preserved by the quantization. From this point of view it is
instructive to investigate the behavior of (2.3-5) when the system is
restricted on $\hl $.

\bigskip
\medskip 

\leftline {\bf B. NLS on the half line}
\bigskip 

The relative action, giving rise both to
the equation of motion (1.1) on $\hl $ and the
boundary condition (1.2) is
$$
A\left [\Phi , \bar \Phi \right ] =
$$
$$
\int_{\R} dt \int_{\hl } dx
\left [ i\overline \Phi (t,x) \der_t \Phi (t,x) -
|\der_x \Phi (t,x)|^2 - g |\Phi (t,x)|^4 \right ]
- \eta \int_{\R } dt\, |\Phi (t,0)|^2
\quad .  \neweq
$$
This action is invariant under time translations, which leads to
conservation of the energy
$$
E\left [\Phi , \bar\Phi \right ]
= \int_{\hl } dx \left [
|\der_x \Phi (t,x)|^2 + g |\Phi (t,x)|^4 \right ] + \eta |\Phi (t,0)|^2 \, .
\neweq
$$
Positivity implies $g \geq 0$ and $\eta \geq 0$, which is the case
we are going to analyze below.

The series (2.3), being a solution of the NLS equation on $\R$,
is {\sl a fortiori} a solution when restricted on $\hl $.
In general however, it does not satisfy the boundary condition (1.2).
In this respect, one has the following

\medskip
\noindent {\it Proposition 1: $\Phi(t,x)$ obeys the boundary condition}
(1.2), {\it provided that $\lambda (k)$ satisfies
$$
\lambda (k) = B(k) \lambda (-k) \quad , \neweq
$$
where
$$
B(k) = {k-i\eta \over k+i\eta } \quad . \neweq
$$}
\medskip
{\noindent \it Proof.} Using (2.13), we will show that 
$\Phi^{(n)}(t,x)$ satisfies (1.2) for any $n\geq 0$. For 
$n=0$ the statement is obvious.
So, let us focus on $\Phi^{(n)}(t,x)$ with $n\geq 1$.
Changing variables in Eq.(2.5) according to
$$
k_{2i-1} = p_i \quad , \qquad k_{2j} = - q_j \quad ,
\qquad i=1,\dots,n \, ;\qquad j=0,\dots,n \, , \neweq
$$
one finds
$$
\lim_{x \downarrow 0}\left( \der_x - \eta \right)
\Phi^{(n)} (t,x) =
$$
$$
\int_{\R^{2n+1}} \prod_{j=0}^{2n} {dk_j \over 2\pi}
f^{(n)}(k_0,...,k_{2n})
\overline \lambda (k_1)\cdots \overline \lambda (k_{2n-1})\, \lambda (-k_{2n})
\cdots \lambda (-k_0)\,  \e^{-it\sum_{j=0}^n (-1)^jk_j^2} \, , \neweq
$$
where
$$
f^{(n)}(k_0,...,k_{2n}) =
{\sum_{j=0}^{2n} k_j - i\eta \over
i\,\prod_{j=1}^{2n} (k_j+k_{j-1})} \quad .
\neweq
$$
Using the simple relations
$$
B(k) B(-k) = B(k)\overline B(k) = 1 \quad , \neweq
$$
one concludes that $f^{(n)}$ in Eq.(2.16) can be equivalently replaced 
by its $B$-symmetrized counterpart
$$
f_B^{(n)}(k_0,...,k_{2n}) = 
\sum_{\sigma_0,\dots,\sigma_{2n} \in \{-1,1\}}{1\over 4^n} 
\left(\prod_{j=0}^{2n} {k_j + i \sigma_j \eta \over k_j + i \eta }\right)
{\sum_{j=0}^{2n}\sigma_j k_j - i\eta \over
i\,\prod_{j=1}^{2n} (\sigma_j k_j+\sigma_{j-1} k_{j-1})}
\quad . \neweq
$$
We shall show now that $f_B^{(n)}$ vanishes identically. 
Eq.(2.19) can be given the more convenient form
$$
f_B^{(n)}(k_0,...,k_{2n}) =
{N^{(n)}(k_0,...,k_{2n})\over 4^n i \prod_{j=0}^{2n} ({k_j + i \eta })
\prod_{j=1}^{2n} (k_j^2-k_{j-1}^2) }
$$
where
$$
N^{(n)}(k_0,...,k_{2n}) = \sum_{\sigma_0,\dots,\sigma_{2n} \in \{-1,1\}}
\prod_{j=0}^{2n} ( k_j + i \sigma_j \eta )
\prod_{j=1}^{2n} (\sigma_j k_j-\sigma_{j-1} k_{j-1})
\left (\sum_{j=0}^{2n} \sigma_j k_j - i\eta \right ) \, .
\neweq
$$
The final step is to prove then that the numerator $N^{(n)}$ vanishes.
One way to show the validity of this quite remarkable identity,
is to introduce the auxiliary function
$$
M^{(n)}(k_0,\dots,k_{2n})=
\sum_{\sigma_0,\dots,\sigma_{2n} \in \{-1,1\}} ( \sigma_0 k_0 - i\eta )
\prod_{j=0}^{2n} ( k_j + i \sigma_j  \eta )
\prod_{j=1}^{2n} (\sigma_j k_j- \sigma_{j-1} k_{j-1})
\quad .
\neweq
$$
Now, after some algebra one derives the recurrence relations
$$
N^{(n)}(k_0,\dots,k_{2n}) =
$$
$$
-4 k_0 k_1 (k_1^2+\eta^2)N^{(n-1)}(k_2,\dots,k_{2n}) +
4 k_0 k_1 (k_1^2-k_0^2)M^{(n-1)}(k_2,\dots,k_{2n}) \quad , \neweq
$$
$$
M^{(n)}(k_0,\dots,k_{2n})  = -4 k_0 k_1 (k_0^2+\eta^2)
M^{(n-1)}(k_2,\dots,k_{2n}) \quad .
\neweq
$$
Since $N^{(0)}(k_0) = M^{(0)}(k_0) = 0$, Eqs.(2.22,23) imply by induction that
$$
N^{(n)}(k_0,\dots,k_{2n}) = 0\quad, \qquad  M^{(n)}(k_0,\dots,k_{2n}) = 0
\quad, \neweq
$$
which completes the argument. 
\medskip 

We conclude here the brief introduction to the classical
boundary value problem (1.1,2). Our next step will be to establish
the quantum counterparts of the solution (2.3-5) and the constraint
(2.13).

\newchapt {III. THE BOUNDARY ALGEBRA}

As already mentioned in the introduction, our basic algebraic tool will
be a particular associative algebra $\balg $, whose generators satisfy specific
quadratic relations.

\medskip 
\bigskip
\leftline {\bf A. Definition of $\balg $}
\bigskip

The concept of boundary algebra has been introduced and investigated in a
general context in Ref. 11. Here we will consider the following special case.
Let $R : \bfR \times \bfR \to \bfC$ be a measurable function
satisfying
$$
R(k_1,k_2)R(k_2,k_1) = R(k_1,k_2) {\overline R}(k_1,k_2)=1
\quad. \neweq
$$
The boundary algebra $\balg $ is
generated by the operator valued distributions\break
$\{a(k),a^*(k),b(k)\, : \, k\in \R \} $,
satisfying quadratic exchange relations,
which can be conveniently grouped in two sets. The first one is
$$
a(k_1) \, a(k_2) - R(k_2,k_1) a(k_2) a(k_1) = 0 \quad , \neweq
$$
$$
a^*(k_1) a^*(k_2) - R(k_2,k_1) a^*(k_2) a^*(k_1) = 0 \quad , \neweq
$$
$$
a(k_1) a^*(k_2) - R(k_1,k_2) a^*(k_2) a(k_1)  =
2\pi \delta (k_1-k_2) + b(k_1)  2\pi \delta (k_1+k_2) \quad . \neweq
$$
The second set of constraints describes
the exchange relations of $b(k)$ and reads
$$
a(k_1) b(k_2) = R(k_2,k_1) R(k_1,-k_2) \,  b(k_2) a(k_1) \quad ,
\neweq
$$
$$
b(k_2) a^*(k_1) = R(k_2,k_1) R(k_1,-k_2) \,  a^*(k_1)b(k_2) \quad ,
\neweq
$$
$$
b(k_1)b(k_2) = b(k_2)b(k_1) \quad . \neweq
$$
Notice that if we formally set $b(k)\to 0$, the relations (3.5-7)
trivialize, while\break (3.2-4) reproduce the defining relations
of the ZF algebra $\alg $. As it is well known, the factorized scattering of
1+1 dimensional integrable systems is encoded in $\alg $, i.e. in
a boundary algebra in which the so called boundary
operator $b(k)$ is trivially implemented. On the contrary,
it turns out$^{11}$ that whenever there is a reflecting boundary,
one needs a {\sl reflection} boundary algebra, i.e. a boundary algebra
with the additional constraint
$$
           b(k) b(-k) = 1
\quad, \neweq
$$
which obviously prevents the boundary operator from being zero.
In the case of the NLS on the half line,
we shall need a reflection boundary algebra $\balg$
with exchange factor
$$
         R(k_1,k_2) = {k_1 - k_2 - i g \over k_1 -k_2 + i g}
\quad,\neweq
$$
where $g\geq 0$ is the coupling constant of the NLS model.
$R(k_1,k_2)$ is actually the two-body bulk scattering matrix
of the NLS model$^{4-9}$ and satisfies (3.1).

\medskip 
\bigskip

\leftline {\bf B. Fock Representations}
\bigskip

{}Following some basic ideas of Ref. 15, we have 
constructed in Ref. 11 the Fock representations of $\balg $. These 
representations are characterized by the existence of a vacuum
state $\Omega $, which is cyclic with respect $a^*(k)$ and 
satisfies 
$$
a(k) \Omega = 0 \quad . \neweq 
$$
In the reflection case (3.8), the vacuum is$^{11}$ always an eigenvector
of the boundary operator $b(k)$, i.e.
$$
b(k) \, \Omega = B(k) \, \Omega
\quad , \neweq
$$
where $B(k)$ is a measurable function obeying Eq.(2.18).
Conversely, any $B(k)$ of this type defines a Fock representation
on a Hilbert space $\brep $, whose vacuum satisfies (3.11). 
We will show below that the state space $\H_{g,\eta }$ of the NLS 
model on $\hl $ is  
$$
\H_{g,\eta } = \brep \quad , \neweq
$$
with $B$ and $R$ given by (2.14) and (3.9) respectively.
The mere fact that our system has a boundary shows up at the algebraic
level, turning the ZF algebra into a reflection boundary algebra
$\balg $, i.e. forcing a non zero boundary operator $b(k)$.
The details of the boundary condition (the value of the parameter $\eta $)
enter at the representation level through the reflection coefficient $B(k)$.
In the Fock space $\brep $ one has 
$$
a(k) = b(k) a(-k) \quad ,
\neweq
$$
$$
a^*(k) = a^*(-k) b(-k)  \quad ,
\neweq
$$
which descend from a peculiar automorphism of $\balg $, established in 
Ref. 11. The relation (3.13) turns out to be the
correct quantum analogue of Eq.(2.13).  Let us stress once more that
the c-number reflection coefficient $B(k)$ must be distinguished from
the boundary generator $b(k)$, which according to Eqs.(3.5,6) does
not even commute with $\{a(k), a^*(k)\}$.

To the end of this section we will give some details about the
structure of $\brep $ which are needed for our construction. One has
$$
\brep \equiv \bigoplus_{n=0}^\infty \H_{R,B}^n \quad , \neweq
$$
where $\H_{R,B}^0 \equiv \C$ and the $n$-particle space $\H_{R,B}^n$
with $n\geq 1$ is a subspace of $L^2(\R^n)$ defined as follows:

\item {(i)} a $L^2$-function $\varphi (p_1)$ belongs to
$\H_{R,B}^1$ if and only if 
$$
\varphi (p_1) = B(p_1) \varphi (-p_1) \quad ; \neweq
$$

\item {(ii)} a $L^2$-function $\varphi (p_1,...,p_n)$ with
$n\geq 2$ belongs to $\H_{R,B}^n$ if and only if
$$
\varphi (p_1,...,p_{n-1},p_n) =
B(p_n) \varphi (p_1,...,p_{n-1},-p_n) \quad , \neweq
$$
and
$$
\varphi (p_1,...,p_i,p_{i+1},...,p_n) = R(p_i,p_{i+1})
\varphi (p_1,...,p_{i+1},p_i,...,p_n) \quad , \neweq
$$
for any $1\leq i\leq n-1$.

\noindent Eqs.(3.16-18) define a closed subspace
$\H_{R,B}^n \subset L^2(\R^n)$. We will denote by $P_{R,B}^{(n)}$
the corresponding orthogonal projection operator. We introduce also
the finite particle space $\F_{R,B}^0 \subset \F_{R,B}$, generated by 
$\{\H_{R,B}^n \, :\, n = 0,1,...\}$. We recall that $\F_{R,B}^0$ is 
the linear space of sequences
$\varphi = \left ( \varphi^{(0)}, \varphi^{(1)},...,\varphi^{(n)},...\right )$
with $\varphi^{(n)}\in \H_{R,B}^n$ and $\varphi^{(n)}=0$ for
$n$ large enough. The vacuum state is $\Omega = (1,0,...,0,...)$.
The $L^2$-scalar product on
$\H^n_{R,B}$ defines in the standard way the scalar product
$\langle \cdot \, ,\, \cdot \rangle $ in the (Hilbert) direct sum (3.15).

At this point we are in position to define on $\F_{R,B}^0$ the
annihilation and creation operators $\{\a , \ac \, :\, f\in L^2(\R ) \}$. 
We set $\a \Omega = 0$ and
$$
\left [\a \varphi \right ]^{(n)}(p_1,...,p_n) =
\sqrt{n+1} \int_{\R} {dp \over 2\pi }\, \overline f(p)
\varphi^{(n+1)} (p, p_1,...,p_n)
\quad , \neweq
$$
$$
\left [\ac \varphi \right ]^{(n)}(p_1,...,p_n)
= \sqrt {n} \left [P^{(n)}_{R,B}\, f\otimes \varphi^{(n-1)}
\right ](p_1,...,p_n) \quad ,
\neweq
$$
{}for all $\varphi \in \F_{R,B}^0$. The operators $\a $ and $\ac $
are in general unbounded on $\F_{R,B}^0$. One can easily see however 
that $\a$ and $\ac $ are bounded on each $\H_{R,B}^n$. In fact, for 
all $\varphi \in \H_{R,B}^n$ one has the estimates 
$$
\Vert \a \varphi \Vert \, \leq \, \sqrt{n} \Vert f \Vert 
\Vert \varphi \Vert \, \, , \quad \quad 
\Vert \ac \varphi \Vert \, \leq \, \sqrt{n+1} \Vert f \Vert 
\Vert \varphi \Vert \, \, , \neweq 
$$
$\Vert \cdot \Vert $ being the $L^2$-norm. Notice
also that $\ac $ is linear in $f$, whereas $\a $ is antilinear.
The operator-valued distributions $a(p)$ and $a^\ast (p)$,
generating the Fock representation of $\balg $, are defined by
$$
\a = \int_{\R } {dp \over 2\pi }\overline f(p)a(p) \quad , \quad \quad
\ac = \int_{\R } {dp \over 2\pi }f(p)a^\ast (p) \quad , \neweq
$$
and are related by Hermitian conjugation, namely
$$
\langle \varphi , \a \psi \rangle = \langle \ac \varphi ,\psi \rangle
\, , \qquad \forall \, \varphi , \psi \in \F_{R,B}^0 \, . \neweq
$$
{}Finally, the action of the boundary generator $b(p)$ on $\F_{R,B}^0$ is
defined by Eq.(3.11) and
$$
\left [b(p) \varphi \right ]^{(n)}(p_1,...,p_n) =
$$
$$
[R(p,p_1) R(p,p_2)\cdots R(p,p_n) B(p)
R(p_n,-p)\cdots R(p_2,-p) R(p_1,-p)]
\varphi^{(n)}(p_1,...,p_n) \quad . \neweq
$$
One can show$^{11}$ that $\{a(p), a^\ast (p), b(p)\}$, defined above,
indeed satisfy the exchange relations (3.2-7) and the reflection condition 
(3.8). Moreover, the vacuum $\Omega $
obeys the requirements formulated in the beginning of this
subsection.

It is convenient to introduce here a domain
$\D \subset \F_{R,B}$, which will be frequently 
used in what follows. Setting  
$$
\D^0 \equiv \C \, , \quad 
\D^n \equiv \left \{ \int_{\R^n} dp_1...dp_n f(p_1,...,p_n)a^*(p_1)...a^*(p_n)\Omega
\, :\, f\in {\cal S}(\R^n), \, n \geq 1 \right \}\, , \neweq 
$$
we define $\D$ to be the linear space of sequences
$\varphi = \left ( \varphi^{(0)}, \varphi^{(1)},...,\varphi^{(n)},...\right )$, 
where\break $\varphi^{(n)}\in \D^n$ and $\varphi^{(n)}$ vanish for $n$ large enough. 
By construction $\D$ is a proper subspace of $\F_{R,B}^0$. Nevertheless, 
$\D$ is dense in $\F_{R,B}$ as well. Indeed, 
using that the factors $R$ and $B$ are smooth (i.e. $C^\infty $) bounded
functions, one has that $\D^n$ is dense in $\H_{R,B}^n$, which 
implies the statement. We observe that 
$$
\a \D \subset \D \, , \quad \ac \D \subset \D \, , 
\quad \forall \, f \in \S(\R ) \quad . \neweq 
$$
Notice also that the matrix elements
of $a^\ast (k)$ between states from $\D$ are
smooth functions of $k$. More generally, one has 
$$
\langle \varphi\, ,\, a^*(k_1)\cdots a^*(k_n)\psi \rangle
\in \S(\R^n )\, , \quad  \forall \, \varphi , \psi \in \D \, . \neweq
$$

Summarizing, we introduced in this section the boundary algebra $\balg $
and its Fock representation $\brep $, which are the main ingredients
in the construction of the quantum solution of the boundary value 
problem (1.1,2).

\newchapt {IV. QUANTIZATION}

\leftline {\bf A. The quantum field $\Phi(t,x)$}

\bigskip

Our first step will be to introduce the quantum analog of
$\Phi^{(n)}(t,x)$. For this purpose we consider
$$
\Phi^{(0)}(t,x) \equiv \widetilde a(t,x) = \int_{\R} {dq \over 2\pi}
a(q)\, \e^{ixq - itq^2} \quad , \neweq
$$
$$
\Phi^{(n)}(t,x) =
$$
$$
\int_{\R^{2n+1} } \prod_{i=1\atop j=0}^n {dp_i \over 2\pi}
{dq_j \over 2\pi}
a^*(p_1)\cdots a^*(p_n)
a(q_n)\cdots a(q_0)
{\e^{i\sum_{j=0}^n (xq_j - tq_j^2 ) - i\sum_{i=1}^n (xp_i - tp_i^2) }
\over \prod_{i=1}^n \left[ (p_i - q_{i-1}-i\epsilon )\,
 (p_i - q_{i}-i\epsilon ) \right]} \, , \neweq
$$
thus replacing formally $\{\lambda (p), \overline \lambda (p)\}$ 
in Eqs.(2.4,5) by
the generators $\{a(p), a^\ast(p)\}$ of $\balg $ in the Fock
representation $\brep $ and fixing an $i\epsilon $ prescription to
contour poles. 
Our first task will be to give meaning of $\Phi^{(n)}(t,x)$ 
as a quadratic form in $\D$:

\medskip
\noindent {\it Proposition 2: For any $\varphi , \psi \in \D$, the
expectation value 
$$
\langle \varphi , \Phi^{(n)}(t,x) \psi \rangle \quad, \neweq
$$
is a $C^\infty $ function of $t,x$. 
\medskip}

\noindent {\it Proof.} The case $n=0$ is trivial. For $n\geq 1$
it is enough to take $\varphi \in \D^m$ and $\psi \in \D^{m+1}$
with $m>n$. Some elementary algebra leads to
$$
\langle \varphi, \Phi^{(n)}(t,x) \psi \rangle =
\int_{\R^{m+n+1} } \prod_{i_1=1}^n {dp_{i_1} \over 2\pi}
\prod_{i_2=0}^n {dq_{i_2} \over 2\pi}
\prod_{i_3=n+1}^m {dk_{i_3} \over 2\pi}
$$
$$
{\overline \varphi }(p_1,...,p_n,k_{n+1},...,k_m)
{\e^{i\sum_{j=0}^n (xq_j - tq_j^2 ) - i\sum_{i=1}^n (xp_i - tp_i^2) }
\over \prod_{i=1}^n \left[ (p_i - q_{i-1} - i \epsilon)\,
(p_i - q_{i} - i \epsilon) \right]}
\psi (q_0,...,q_n,k_{n+1},...,k_m) \, , \neweq
$$
which, using that $\varphi $ and $\psi $ are Schwartz test functions,
implies the proposition.
\medskip 

Taking into account that $\D$ contains only finite particle vectors, 
we conclude that also $\Phi(t,x)$ is a quadratic form on $\D$, 
smooth in both $t$ and $x$. The conjugate $\Phi^*(t,x)$ 
is defined by 
$$
\langle \varphi , \Phi^*(t,x) \psi \rangle  =  
{\overline {\langle \psi , \Phi(t,x) \varphi \rangle }}
\, ,\neweq
$$
which is of course smooth in $t$ and $x$ as well. The counterparts 
of Eqs.(4.1,2) read 
$$
\Phi^{*(0)}(t,x) \equiv \widetilde a^*(t,x) = \int_{\R} {dq \over 2\pi} 
a^*(q)\, \e^{-ixq + itq^2} \quad , \neweq 
$$
$$
\Phi^{*(n)}(t,x) = 
$$
$$
\int_{\R^{2n+1} } \prod_{i=1\atop j=0}^n {dp_i \over 2\pi}
{dq_j \over 2\pi}
a^*(q_0)\cdots a^*(q_n) 
a(p_n)\cdots a(p_1) 
{\e^{ i\sum_{i=1}^n (xp_i - tp_i^2) -i\sum_{j=0}^n (xq_j - tq_j^2 ) } 
\over \prod_{i=1}^n \left[ (p_i - q_{i-1}+i\epsilon )\,
 (p_i - q_{i}+i\epsilon ) \right]} \, . \neweq  
$$ 

Since the system we are considering is in $\hl $, we adopt the
smearing 
$$
\Phi(t,f) = \int dx {\overline f}(x) \Phi(t,x) \, , \qquad
\Phi^*(t,f) = \int dx {f}(x) \Phi^*(t,x) \, ,
\quad f\in C^\infty_0(\hl ) \quad , 
\neweq
$$
where $C^\infty_0(\hl )$ is the set of infinitely differentiable functions 
with compact support in $\hl$. Again, $\Phi(t,f)$ and $\Phi^*(t,f)$ have meaning 
as quadratic forms on $\D$, which are related by 
$$
    \langle \varphi , \Phi^*(t,f) \psi \rangle  =  
    \overline{\langle \psi , \Phi(t,f) \varphi \rangle }
\, .\neweq
$$

In order to formulate some other less obvious properties of 
$\Phi(t,f)$ and $\Phi^*(t,f)$, we have to introduce 
the following partial ordering relation in $C^\infty_0(\hl )$. Let\break 
$f_1,\, f_2\in C^\infty_0(\hl)$. Then
$$
f_1\prec f_2 \quad  \Longleftrightarrow \quad x_1<x_2\quad
\forall \, x_1\in\supp f_1 \, , \quad \forall \, x_2\in\supp f_2
\quad. \neweq
$$
Instead of $f_1 \prec f_2$, we will also write $f_2 \succ f_1$. 
Denoting by $\at^*(t,f)$ the operator 
$$
\at^*(t,f) = \int dx {f}(x) \at^*(t,x) \quad , \neweq
$$
one can prove the following technical 

\medskip

\noindent {\it Lemma 1: Let $\varphi,\psi\in\D$. 

\noindent (a) The identity
$$
\langle \varphi , \Phi^*(t,h) \at^*(t,f) \psi \rangle = 
\langle \varphi ,\at^*(t,f) \Phi^*(t,h) \psi \rangle
\quad, \neweq
$$
holds if $h \prec f$;

\noindent (b) One has
$$
\langle \varphi \, ,\, \Phi^*(t,h)\at^*(t,f_1)\cdots \at^*(t,f_n)\Omega \rangle=
\langle \varphi , \at^*(t,h)\at^*(t,f_1)\cdots \at^*(t,f_n)\Omega \rangle\quad , \neweq
$$
provided that $h \succ f_j$ for any $j=1,...,n$;

\noindent (c) For any  
$f_1\succ f_2\succ ...\succ f_n$, one has
$$
\langle \varphi \, ,\, \Phi (t,h)\at^*(t,f_1)\at^*(t,f_2)\cdots \at^*(t,f_n)
\Omega \rangle =
$$
$$
\sum_{j=1}^n (h\, , f_j) 
\langle \varphi \, ,\, \at^*(t,f_1)\cdots \widehat {\at^*} (t,f_j) 
\cdots \at^*(t,f_n)\Omega \rangle
\, , \neweq
$$
where $(\cdot \, ,\, \cdot )$ denotes the $L^2$-scalar product 
and the hat indicates that the corresponding field must be omitted.}
\medskip

\noindent {\it Proof.} The proof of the identities (4.12-14) is
analogous to that given by Davies$^8$ for the NLS on $\R $, so we 
skip it. We only remark that 
the novelty on $\hl $ consists in evaluating the contributions of the
boundary generator $b$, which stem from the exchange of $a$ and $a^*$.
It is easy to see that these contributions actually vanish,
due to the support requirements imposed on the test functions
and the condition $\eta \geq 0$.
\medskip

Summarizing, $\Phi (t,f)$ and $\Phi^*(t,f)$ have been so far 
defined as quadratic forms on $\D$ and are Schwartz 
distributions with respect to $f$. Our main goal to the end of this 
subsection will be to show that $\Phi (t,f)$ and $\Phi^*(t,f)$ are 
actually well defined operators. In order to construct 
a common invariant domain for these operators, we introduce the subspace   
$$
\D_0^n \equiv \sp \, \{ \at^*(t,f_1)\at^*(t,f_2)\cdots\at^*(t,f_n)\Omega
\, : \, f_1 \succ f_2 \succ \cdots \succ f_n \}\subset \H_{R,B}^n \, , 
\quad  n\geq 1 \, , \neweq
$$
where $\sp$ indicates the linear span and $t\in \R$ is arbitrary but fixed. 
Setting $\D_0^0 = \C$, we define $\D_0$ to be the linear space of sequences
$\varphi = \left ( \varphi^{(0)}, \varphi^{(1)},...,\varphi^{(n)},...\right )$
with $\varphi^{(n)}\in \D_0^n$ and $\varphi^{(n)}=0$ for $n$ large enough. 
Both $\D$ and $\D_0$ are subspaces of the finite particle space $\F_{R,B}^0$. 
We know already that $\D$ is dense in $\F_{R,B}$. Although it is less obvious, 
the same is true for $\D_0$. 

\medskip 

\noindent {\it Proposition 3: $\D_0$ is is dense in $\F_{R,B}$.}
\medskip

{\noindent \it Proof.} 
It is enough to demonstrate that the space $\D_0^n $
is dense in $\H^n_{R,B}$ for any $t\in \R$ and $n\geq 1$. So,
let us consider the matrix element
$$
\widetilde A_{t,\varphi }(x_1,...,x_n) \equiv
\langle \varphi\, ,\, \at^*(t,x_1)\cdots\at^*(t,x_n) \Omega
\rangle \quad , \neweq
$$
where $\varphi \in \D^n$ is arbitrary. 
According to Eq.(3.27) $\widetilde A_{t, \varphi }\in \S(\R^n)$.
In order to prove the statement, it is sufficient to show that
$$
\widetilde A_{t, \varphi }(x_1,...,x_n) = 0 \, , \quad
\forall \, x_1 > x_2 > ... >x_n > 0 \, , \neweq
$$
implies $\varphi = 0$. It is convenient for
this purpose to investigate
$$
A_{t, \varphi }(p_1,...,p_n) \equiv 
$$
$$
\int_{\R^n }\prod_{j=1}^n dx_j \, \e^{i\sum_{j=1}^np_jx_j }
\widetilde A_{t, \varphi }(x_1,...,x_n) =
\e^{it\sum_{j=1}^n p^2_j}
\langle \varphi\, ,\, a^*(p_1)\cdots a^*(p_n) \Omega
\rangle \in \S(\R^n ) \, . \neweq
$$
The behavior of this function under the reflection of one of its
arguments or the exchange of two consecutive arguments is determined
by Eqs.(3.3,6,11,14). Using this fact, one can verify that the function
$$
B_{t, \varphi }(p_1,...,p_n) \equiv \Lambda (p_1,...,p_n)
A_{t, \varphi }(p_1,...,p_n) \quad , \neweq
$$
where
$$
\Lambda (p_1,...,p_n) \equiv
\prod_{j=1}^n \left [(p_j-i\eta)
\prod_{k=1\atop k>j}^n (p_j-p_k -ig)(p_j+p_k -ig)\right ]
\neweq
$$
satisfies
$$
B_{t, \varphi }(p_1,...,p_j,...,p_n)
= - B_{t, \varphi }(p_1,...,-p_j,...,p_n) \, , \quad
\forall \, j = 1,...,n \, , \neweq
$$
$$
B_{t, \varphi }(p_1,...,p_j,p_{j+1},...,p_n)
= - B_{t, \varphi }(p_1,...,p_{j+1},p_j,...,p_n) \, , \quad
\forall \, j = 1,...,n-1 \, . \neweq
$$
By construction $B_{t, \varphi }\in \S(\R^n )$ and
$$
\widetilde B_{t, \varphi }(x_1,...,x_n) =
$$
$$
\int_{\R^n} \prod_{j=1}^n {dp_j \over 2\pi}
\, \e^{-i\sum_{j=1}^np_jx_j } B_{t, \varphi }(p_1,...,p_n)
= \Lambda (i\der_1,...,i\der_n )
\widetilde A_{t, \varphi }(x_1,...,x_n) \quad , \neweq
$$
admits the same antisymmetry properties as $B_{t, \varphi }$.
Therefore, using the smoothness of $\widetilde A_{t, \varphi }$ and
Eq.(4.17), we deduce that $\widetilde B_{t, \varphi }$ vanishes 
identically, or equivalently,
$$
B_{t, \varphi }(p_1,...,p_n) = 0 \, , \quad \forall \, p_j \in \R \, .
\neweq
$$
Combining Eqs.(4.18,19,24) with the fact that
$\Lambda (p_1,...,p_n) \not= 0$ for any $p_j\in \R$, one gets
$$
\langle \varphi\, ,\, a^*(p_1)\cdots a^*(p_n) \Omega \rangle = 0
\, , \quad \forall \, p_j \in \R \, , \neweq
$$
which, because of the cyclicity of $\Omega $ with respect to
$a^*$, implies $\varphi = 0$. This concludes the argument.
\medskip

\noindent It is convenient in what follows to have an explicit formula for 
the scalar product in $\D_0$. It is provided by the following

\medskip
\noindent{  \it Lemma 2}: {\it Let $f_1 \succ  f_2 \succ \cdots \succ f_n$ and 
$h_1 \succ  h_2 \succ \cdots \succ h_n$. Then}  
$$
   \langle \at^*(t,h_1) \cdots \at^*(t,h_n) \Omega \, , \,
           \at^*(t,f_1) \cdots \at^*(t,f_n) \Omega \rangle
= (h_1 \otimes \cdots \otimes h_n \, , \, f_1 \otimes \cdots \otimes f_n)
\quad. \neweq
$$
\medskip
{\noindent \it Proof.} It is enough to expand the left hand side, 
using the algebraic relations (3.4) and Eq.(3.10). Taking into 
account the support properties of the test functions involved, 
all terms, except the one in the right hand side of (4.26), vanish. 
\medskip

A simple corollary of the previous lemma is now in order. Since 
any $\varphi \in\D_0^n$ can be represented as
$$
\varphi = 
\sum_{\alpha \in A} \at^*(t, f^\alpha_1) \cdots \at^*(t, f^\alpha _n) \Omega 
\quad , 
$$
where $A$ is a finite set and 
$f^\alpha_1 \succ  f^\alpha_2 \succ \cdots \succ f^\alpha_n$ for 
all $\alpha \in A$, one has that 
$$
\langle \varphi \, ,\, \varphi \rangle^2 \equiv 
\Vert \varphi \Vert^2  = 
\Vert \sum_{\alpha \in A} f^\alpha_1 \otimes \cdots \otimes f^\alpha_n \Vert^2  
\quad. \neweq
$$

We are now in position to show the following 

\noindent {\it Proposition 4: The estimate 
$$
|\langle \varphi , \Phi(t,f) \psi \rangle | \le (n+1)
\Vert f \Vert \, \Vert \varphi \Vert \, \Vert \psi \Vert
\neweq
$$ 
holds for any $\varphi\in \D_0^n$,  $\psi \in \D_0^{n+1}$ 
and $f\in C^\infty_0(\hl)$.} 
\medskip

{\noindent \it Proof.} 
Let 
$$
\varphi = 
\sum_{\alpha \in A} \at^*(t, f^\alpha_1) \cdots \at^*(t, f^\alpha _n) \Omega \, , \qquad
\psi = 
\sum_{\beta \in B} \at^*(t, h^\beta_0) \cdots \at^*(t, h^\beta _n) \Omega \, ,\neweq
$$
with $f^\alpha_1 \succ  f^\alpha_2 \succ \cdots \succ f^\alpha_n$ and 
$h^\beta_0 \succ  h^\beta_2 \succ \cdots \succ h^\beta_n$. 
Then 
$$
\langle \varphi , \Phi(t,f) \psi \rangle  = 
\sum_{\alpha \in A} \sum_{\beta \in B} 
\langle   \at^*(t, f^\alpha_1) \cdots \at^*(t, f^\alpha _n) \Omega \, , \,
\Phi(t,f) \, \at^*(t, h^\beta_0) \cdots \at^*(t, h^\beta _n) \Omega\rangle 
$$
$$
= \sum_{\alpha \in A} \sum_{\beta \in B} \sum_{j=0}^n (f,h_j^\alpha) 
(f_1^\alpha \otimes \cdots \otimes f_n^\alpha \, , \, h_0^\beta \otimes \cdots 
\otimes {\widehat h}_j^\beta \otimes \cdots \otimes h_n^\beta ) 
$$
$$
= \sum_{j=0}^n 
( \sum_{\alpha \in A} f_1^\alpha \otimes \cdots f_{j-1}^\alpha \otimes f \otimes f_j^\alpha 
\cdots \otimes f_n^\alpha \, ,\, \sum_{\beta \in B}  h_0^\beta \otimes 
\cdots \otimes {h}_j^\beta \otimes \cdots \otimes  h_n^\beta ) \, , 
$$
where use has been made of point (c) of Lemma 1.
Applying now the Minkowski inequality, one finds 
$$
|\langle \varphi , \Phi(t,f) \psi \rangle | \le \sum_{j=0}^n \Vert f \Vert
\Vert \sum_{\alpha \in A} f_1^\alpha \otimes 
\cdots \otimes f_n^\alpha \Vert 
\Vert \sum_{\beta \in B} h_0^\beta \otimes 
\cdots \otimes  h_n^\beta \Vert 
\le (n+1)
\Vert f \Vert \, \Vert \varphi \Vert \, \Vert \psi \Vert
\, . \neweq
$$

The above proposition shows that $\Phi(t,f)$, considered as 
quadratic form, is bounded on $\D_0^n\times\D_0^{n+1}$ and defines therefore 
a bounded operator $\H_{R,B}^{n+1} \to \H_{R,B}^{n}$. Since this occurs for 
any $n \geq 0$, we recover an operator $\Phi(t,f)\, :\, \F_{R,B}^0\to\F_{R,B}^0$, 
whose properties are collected in 

\medskip
\noindent {\bf Theorem 1:} {\it $\Phi (t,f)\, :\, \F_{R,B}^0\to\F_{R,B}^0$ 
is a linear operator, satisfying 
$$
\Phi(t,f) \Omega = 0 \, , 
\qquad \Phi (t,f)\, :\, \H_{R,B}^{n+1} \to \H_{R,B}^n \, , 
\qquad n\geq 0 \, . \neweq
$$ 
Moreover, for any $\varphi,\psi\in \F_{R,B}^0 $, 
the matrix element $\langle \varphi ,\Phi (t,f) \psi \rangle $
has the following properties:
\item{i)} {It is antilinear and $L^2$-continuous in $f$;} 
\item{ii)} {It is continuous in $t\in\R$;}
\item{iii)} {It is smooth in $t\in \R$, provided that $\varphi , \psi \in \D$.}}
\medskip
\noindent{\it Proof:} All the statements are simple corollaries of the
above propositions. 
\medskip

The operator $\Phi (t,f)$ is densely defined and admits therefore a
Hermitian conjugate $\Phi^* (t,f)$. 
\medskip
\noindent {\bf Theorem 2} {\it : The field $\Phi^* (t,f)$ satisfies
$$
\Phi^* (t,f)\Omega = \at^*(t,f)\Omega \, , \qquad 
\Phi^* (t,f)\, :\, \H_{R,B}^n \to \H_{R,B}^{n+1} \, , \qquad n\geq 0
\neweq
$$
and therefore leaves $\F_{R,B}^0$ invariant. Moreover
$$
\langle \varphi , \Phi (t,f) \psi \rangle =
\langle \Phi^* (t,f) \varphi , \psi \rangle \quad , \neweq
$$
holds for any $\varphi , \psi \in \F_{R,B}^0$.}
\medskip

\noindent{\it Proof:} One uses the fact that $\Phi(t,f)$ is
bounded on each $\H_{R,B}^n$. 

\medskip 

We will show now that the operators $\Phi(t,f)$ and $\Phi^*(t,f)$ 
satisfy the basic requirements for nonrelativistic quantum fields. 

\medskip 
\bigskip 

\leftline {\bf B. Cyclicity of $\Omega $ and commutation relations}

\bigskip 

We start with 

\medskip
{\noindent \bf Theorem 3} (Cyclicity) {\it : The vacuum $\Omega$ is a cyclic vector 
for the field $\Phi^*$. More precisely the space 
$$
\E_0^n \equiv \sp \, \{ \Phi^*(t,f_1)\Phi^*(t,f_2)\cdots\Phi^*(t,f_n)\Omega
\, : \, f_1 \prec f_2 \prec \cdots \prec f_n \}\, , 
$$
is dense in $\H_{R,B}^n$.}
\medskip

\noindent {\it Proof.} Using Eqs.(4.12-13) of Lemma 1, one easily 
proves by induction that 
$$
\Phi^*(t,f_1)\Phi^*(t,f_2)\cdots\Phi^*(t,f_n)\Omega =
\at^*(t,f_n)\cdots \at^*(t,f_1)\Omega \, , \neweq 
$$ 
as long as $f_1 \prec f_2 \prec \cdots \prec f_n $. Thus 
$\E_0^n = \D_0^n$, and the statement follows directly from 
Proposition 3.
\medskip

\noindent {\it Remark:} Theorem 3 is slightly stronger then 
the standard cyclicity$^{16}$, because of the ordering
among the functions $f_1,\ldots,f_n$ required in the definition of $\E_0^n$.
\medskip
\noindent Let us consider now the canonical commutation relations (1.3,4). 
We shall prove 

\noindent {\bf Theorem 4} : {\it The equal time canonical
commutation relations
$$
[\Phi (t,h_1)\, ,\, \Phi (t,h_2)] =
[\Phi^* (t,h_1)\, ,\, \Phi^* (t,h_2)] = 0 \quad , \neweq
$$
$$
[\Phi (t,h_1)\, ,\, \Phi^* (t,h_2)] =
(h_1\, , h_2) \quad , \neweq
$$
hold on $\F_{R,B}^0$ for any $h_1,h_2 \in \S(\hl )$.}
\medskip

\noindent {\it Proof:} In order to demonstrate Eq.(4.35), we observe
that Eq.(4.14) implies
$$
\Phi (t,h_2)\at^*(t,f_1)\cdots \at^*(t,f_n)\Omega =
\sum_{j=1}^n (h_2\, , f_j)
\at^*(t,f_1)\cdots \widehat {\at^*}(t,f_j) \cdots \at^*(t,f_n)\Omega
\, , 
$$
where $f_1 \succ ...\succ f_n$. Therefore
$$
\Phi (t,h_1)\Phi (t,h_2)
\at^*(t,f_1)\cdots \at^*(t,f_n)\Omega =
$$
$$
\sum_{j=1}^n \sum_{k=1\atop k\not=j}^n  (h_2\, , f_j) (h_1\, , f_k)
\at^*(t,f_1)\cdots \widehat {\at^*}(t,f_j) \cdots
\widehat {\at^*}(t,f_k) \cdots \at^*(t,f_n)\Omega \, , \neweq 
$$
which, being symmetric under the exchange of $h_1$ with $h_2$,
implies the vanishing of $[\Phi (t,h_1)\, ,\, \Phi (t,h_2)]$ on
$\D_0^n$. Then one extends by continuity to $\H_{R,B}^n$ and by
linearity to $\F_{R,B}^0$.
The validity of $[\Phi^* (t,h_1)\, ,\, \Phi^* (t,h_2)] = 0$
follows by Hermitian conjugation.
 
We turn now to Eq.(4.36). Let $f_1 \succ ...\succ f_n$ 
and $h_1,h_2\in \S(\hl)$. Assume that
$$
f_k \succ h_2 \succ f_{k+1} \quad . 
\neweq 
$$
Using Lemma 1, one gets
$$
\Phi (t,h_1)\Phi^* (t,h_2)
\at^*(t,f_1)\cdots \at^*(t,f_n)\Omega  =
$$
$$
 \Phi (t,h_1)
\at^*(t,f_1)\cdots \at^*(t,f_k)\Phi^* (t,h_2) \at^*(t,f_{k+1})\cdots
\at^*(t,f_n)\Omega  =
$$
$$
 \Phi (t,h_1)
\at^*(t,f_1)\cdots \at^*(t,f_k)\at^* (t,h_2) \at^*(t,f_{k+1})\cdots
\at^*(t,f_n)\Omega  =
$$
$$
= (h_1,h_2)  \,  \at^*(t,f_1)
\cdots \at^*(t,f_n)\Omega  +
$$
$$
\sum_{j=1}^n (h_1\, , f_j) \, \at^*(t,f_1)
\cdots \widehat {\at^*}(t,f_j)
\cdots \at^*(t,f_k)\at^* (t,h_2) \at^*(t,f_{k+1})\cdots
\at^*(t,f_n)\Omega 
\quad . 
$$
Analogously,
$$
 \Phi^* (t,h_2) \Phi (t,h_1)
\at^*(t,f_1)\cdots \at^*(t,f_n)\Omega =
$$
$$
\sum_{j=1}^n (h_1\, , f_j) \, \at^*(t,f_1)
\cdots \widehat {\at^*}(t,f_j)
\cdots \at^*(t,f_k)\at^* (t,h_2) \at^*(t,f_{k+1})\cdots
\at^*(t,f_n)\Omega 
\quad . 
$$
Therefore
$$
[\Phi (t,h_1)\, ,\, \Phi^* (t,h_2)]
\at^*(t,f_1)\cdots \at^*(t,f_n)\Omega 
= (h_1,h_2)  \at^*(t,f_1)\cdots
\at^*(t,f_n)\Omega 
\quad . \neweq
$$
So, Eq.(4.36) holds on states of the type 
$\at^*(t,f_1)\cdots \at^*(t,f_n)\Omega $, which satisfy the condition 
(4.38). Observing that the couples $\{h_2\, , \at^*(t,f_1)\cdots \at^*(t,f_n)\Omega \}$ 
obeying (4.38) are norm dense in $L^2(\hl) \otimes \H^n_{RB}$, Eq.(4.36) follows 
by continuity.   

\medskip

As a consequence of the commutation relations (4.35,36), 
one has the following useful estimate. 

\medskip 
\noindent {\it Proposition 5: Let $A$ be a finite set 
and let $f^\alpha_1 , ... , f^\alpha_n \in C^\infty_0$ for any 
$\alpha \in A$. Then the norm of the operator 
$$
\sum_{\alpha \in A} \Phi(t,f^\alpha_1)\Phi(t,f^\alpha_2)
\cdots \Phi(t,f^\alpha_n) \, , 
$$
restricted to $\H^m_{R,B}$ with $m\ge n$, satisfies 
$$
\Vert \sum_{\alpha \in A} 
\Phi(t,f^\alpha_1)\Phi(t,f^\alpha_2)\cdots \Phi(t,f^\alpha_n) \Vert
\le  \sqrt{ m(m-1)\cdots (m-n+1) } \, 
\Vert \sum_{\alpha \in A} f^\alpha_1 \otimes \cdots \otimes f^\alpha_n \Vert 
\quad. \neweq
$$}
\medskip

\noindent {\it Proof: } Let $\psi\in \D_0^n$. Then there is some finite 
set $B$, such that $\psi$ can be written in the form
$$
\psi = \sum_{\beta \in B} \Phi^*(t,h_1^\beta) \cdots \Phi^*(t,h_m^\beta) \Omega
\, , \qquad h_1^\beta \prec \cdots \prec h_m^\beta \, . 
$$
Now, by means of the commutation relations (4.35,36), one finds 
$$
\Vert \sum_{\alpha \in A} \Phi(t,f^\alpha_1)\cdots \Phi(t,f^\alpha_n) \psi \Vert
\leq  \sqrt{ m(m-1)\cdots (m-n+1) } \, 
\Vert \sum_{\alpha \in A} f^\alpha_1 \otimes \cdots \otimes f^\alpha_n \Vert  
\Vert \psi \Vert   \, , \neweq 
$$
implying Eq.(4.40) by continuity. 
\medskip

\newchapt {V. TIME EVOLUTION}

In order to investigate the time evolution in the NLS model 
on $\hl $, we consider the mapping 
$$
\alpha_t( a(k)) = \e^{-i k^2 t} a(k) \, , \quad 
\alpha_t( a^*(k)) = \e^{i k^2 t} a^*(k) \, ,\quad  
\alpha_t( b(k)) = b(k)  \, , \quad t\in \R \, .
\neweq
$$
It is straightforward to verify that $\alpha_t$  
defines a 1-parameter group of automorphisms of the boundary algebra 
$\balg $. Using the relations (3.2-6,13,14), one can easily check 
that this group is unitarily implemented in the Fock space $\F_{R,B}$ 
by means of the operator  
$$
U(t)=\exp(i H t) \, , \qquad 
H = {1\over 2} \int_{\R } {dk \over 2\pi} \, k^2 a^*(k) a(k) 
\, . \neweq
$$ 
The Hamiltonian $H$ acts on $\D$ according to 
$$
[H \varphi ]^{(n)}(k_1,...,k_n) = (k_1^2+ \cdots + k_n^2) \,
\varphi^{(n)}(k_1,...,k_n)
\quad ,
\neweq
$$
which implies that the domain $\D$ is invariant both under $U(t)$ and $H$. 
Moreover, since 
$$
-i{d\over dt} U(t)|_{{}_{t=0}} = H
\quad, \neweq
$$
on $\D$, the latter is a domain of essential self-adjointness for $H$. 

The crucial point now is that the time evolution of the field 
$\Phi(t,f)$ is given by
$$ 
\Phi(t,f) = U(t)\, \Phi(0,f) \, U(t)^{-1} \quad .\neweq
$$
This fact follows directly from the time dependence encoded 
in Eqs.(4.1,2) and is quite remarkable. It shows the power of 
both the quantum inverse scattering transform (4.2) and 
the algebra $\balg $, which combined together 
allow to write down the Hamiltonian of an interacting 
field theory as a simple quadratic expression in $a$ and $a^*$. 
In this form $H$ depends only implicitly on the coupling 
constant $g$ through the exchange factor $R$. Notice also 
that the boundary generator $b$ does not evolve in time. 
\bigskip 
\medskip 
\leftline {\bf A. The quantum equation of motion} 

\bigskip 

A preliminary problem to be faced here is to give a precise meaning 
on the quantum level of the cubic term $|\Phi (t,x)|^2\Phi (t,x)$ present 
in Eq.(1.1). For this purpose we will follow the 
standard approach, introducing the concept of a normal ordered $: ... :$ 
product involving $\Phi $ and $\Phi^*$. As usually assumed, 
in such a product all creation operators $a^*$ stand to the left of 
all annihilation operators $a$. In view of Eqs.(3.2,3), in our case one 
must further specify the ordering of creators and annihilators 
themselves. We define $: ... :$ to preserve the original order 
of the creators. The original order of two annihilators is 
preserved if both belong to the same $\Phi $ or $\Phi^*$ 
and inverted otherwise. The quantum version of Eq.(1.1) 
is then obtained by the substitution
$$
|\Phi (t,x)|^2 \Phi (t,x)\, \mapsto \; : \Phi \Phi^* \Phi : (t,x)
\quad . \neweq 
$$
Concerning the relation between the above way of defining 
the normal product and the alternative point-splitting procedure, 
we observe that  
$$
:\Phi\Phi^*\Phi: (t,x) =
\lim_{\sigma \downarrow 0}
\Phi(t,x+2\sigma ) \Phi^*(t,x+\sigma ) \Phi (t,x)
\quad , \neweq
$$
holds in mean value on $\D$. Following Ref. 6, Eq.(5.7) can be derived 
by using the analyticity properties of the commutator between 
$a(p)$ and $\Phi (t,x)$. One can formulate at this point  

\noindent {\bf Theorem 5:} {\it The Nonlinear Schr\"odinger equation 
$$
(i\der_t + \der_x^2 )\langle \varphi \, ,\, \Phi (t,x) \psi \rangle  
= 2g\, \langle \varphi \, ,\, :\Phi \Phi^* \Phi: (t,x) \psi \rangle 
\quad ,
\neweq
$$
is satisfied for any $\varphi ,\, \psi \in \D$.} 
\medskip 

\noindent {\it Proof:} The first step is analogous to the proof 
of Proposition 2 and consists in showing that the matrix element 
$\langle \varphi \, ,\,  : \Phi \Phi^* \Phi : (t,x) \psi \rangle $ 
is smooth in $t$ and $x$ for any $\varphi ,\psi \in \D$. 
The next step is to compare 
$(i\der_t + \der_x^2 )\langle \varphi \, ,\Phi^{(n)} (t,x)\psi \rangle $ 
with the $(n-1)$-th order term in the expansion of 
$\langle \varphi \, ,\,  : \Phi \Phi^* \Phi : (t,x) \psi \rangle $ 
in terms of $g$. A straightforward computation, similar to 
that performed in Ref. 8 for the NLS model on $\R$, shows that 
these terms indeed coincide. 

\medskip 
\bigskip

\leftline {\bf B. Boundary conditions}

\bigskip
We shall demonstrate now 

\medskip
\noindent {\bf Theorem 6.} {\it The following boundary
conditions hold for any $\varphi ,\, \psi \in \D $ and $t\in \R$}: 
$$
\lim_{x \downarrow 0}
\left ( \der_x - \eta \right )
\langle \varphi \, ,\, \Phi(t, x) \psi \rangle = 0
\, , \neweq
$$
$$
\lim_{x\to\infty }
\langle \varphi \, ,\, \Phi(t, x) \psi \rangle = 0 \, . \qquad
\neweq
$$

\medskip

\noindent Let us first prove 

\medskip
\noindent{  \it Lemma 3}: {\it Let $\varphi,\psi\in\F^0_{R,B}$. There 
exists a vector $\chi\in \H^1_{R,B}$ such that}
$$
     \langle \varphi \, ,\, \Phi(t, f) \psi \rangle = 
\langle \Omega \, ,\, \Phi(t, f) \chi \rangle 
\quad. \neweq
$$
\medskip

\noindent {\it Proof:} Without loss of generality one can take 
$\varphi\in\H^n_{R,B}$, $\psi\in\H^{n+1}_{R,B}$. Suppose first
that $\varphi\in\E_0^n = \D_0^n$. Then $\varphi $ is of the form 
$$
\varphi=\sum_{\alpha\in A} 
\Phi^*(t,f^\alpha_1)\Phi^*(t,f^\alpha_2)\cdots\Phi^*(t,f^\alpha_n)\Omega
\quad, 
\neweq
$$
where $A$ is a finite set and 
$ f^\alpha_1 \prec f^\alpha_2 \prec \cdots \prec f^\alpha_n$
for all $\alpha\in A$. 
Using the commutation relations (4.35,36), one easily obtains
$$
     \langle \varphi \, ,\, \Phi(t, f) \psi \rangle = 
\sum_{\alpha\in A}   \langle \Omega \, ,\, \Phi(t, f) \, 
\Phi(t,f^\alpha_n)\Phi(t,f^\alpha_{n-1})\cdots\Phi(t,f^\alpha_1)
\psi \rangle
\quad. \neweq
$$
In order to solve (5.11), it is then sufficient to define 
$$
\chi=\Phi(t,f^\alpha_n)\Phi(t,f^\alpha_{n-1})\cdots\Phi(t,f^\alpha_1)\psi 
\quad , \neweq
$$
which belongs to $\H^1_{R,B}$ since $\psi\in\H^{n+1}_{R,B}$.
Take now a general $\varphi\in\H^n_{R,B}$. 
By cyclicity (Theorem 3), there exists a sequence 
$\{\varphi_k \}\subset \D^n_0$ converging to $\varphi$. 
By Proposition 5, the corresponding vectors $\{\chi_k\}$ 
given by Eq.(5.14) form a Cauchy sequence, which converges 
to a vector $\chi\in\H^1_{R,B}$, satisfying (5.11) by continuity.  

\medskip 

We can now prove Theorem 6. 

\medskip

\noindent {\it Proof:} Let $\varphi,\psi\in\D_0\subset\F^0_{R,B}$. 
{}From the lemma above there
exists $\chi\in \H^1_{R,B}$ such that
$$
\langle \varphi \, ,\, \Phi(t, x) \psi \rangle = 
\langle \Omega \, ,\, \Phi(t, x) \chi \rangle =
\int_\R  {dk \over 2\pi} \, \e^{i k x - i k^2 t} \, \chi(k)
\quad. \neweq
$$
Since $\chi \in L^2$, the matrix element 
$\langle \varphi \, ,\, \Phi(t, x) \psi \rangle $, which by Proposition 2
is smooth, is also square integrable with respect to $x$. Therefore it
vanishes at infinity and Eq.(5.10) is satisfied. 
Moreover, taking the derivative with respect to
$x$, the $B$-symmetry (3.16) of $\chi$, immediately leads to Eq.(5.9).  

\bigskip 
\medskip 
\leftline {\bf C. Correlation Functions} 
\bigskip 

{}From the general structure of our solution it follows that: 

\item {(i)} the nonvanishing correlation functions involve
equal number of $\Phi $ and $\Phi^*$; 

\item {(ii)} for computing the exact $2n$-point
function one does not need all terms in the expansion (2.3), 
but at most the $(n-1)$-th order contribution. 

\noindent One has for instance: 
$$
\langle \Omega \, ,\, \Phi(t_1,x_1)\Phi^*(t_2,x_2)\Omega \rangle = 
\langle \Omega \, ,\, \Phi^{(0)}(t_1,x_1)\Phi^{*(0)}(t_2,x_2)\Omega 
\rangle \, , \neweq  
$$
$$
\langle \Omega \, ,\, \Phi(t_1,x_1)\Phi(t_2,x_2)
\Phi^*(t_3,x_3)\Phi^*(t_4,x_4) \Omega \rangle = 
$$
$$
\langle \Omega \, ,\, \Phi^{(0)}(t_1,x_1)\Phi^{(0)}(t_2,x_2)
\Phi^{*(0)}(t_3,x_3)\Phi^{*(0)}(t_4,x_4) \Omega \rangle + 
$$
$$
g^2 \langle \Omega \, ,\, \Phi^{(0)}(t_1,x_1)\Phi^{(1)}(t_2,x_2)
\Phi^{*(1)}(t_3,x_3)\Phi^{*(0)}(t_4,x_4) \Omega \rangle 
\, . \neweq 
$$
Since the vacuum expectation value of any number of 
$\{a(k), \, a^*(k),\, b(k)\}$ is known explicitly$^{11}$, 
employing Eqs.(4.1,2,6,7) one can derive integral 
representations for the NLS correlation functions on $\hl $. For 
example, 
$$
\langle \Omega \, ,\, \Phi(t_1,x_1)\Phi^*(t_2,x_2)\Omega \rangle = 
\int_\R {dp \over 2\pi} {\rm e}^{-ip^2 (t_1-t_2)}
\left [ {\rm e}^{ip(x_1-x_2)}+B(p) {\rm e}^{ip(x_1+x_2)} \right ] \; , 
\neweq 
$$
which coincides with that of the non-relativistic free field 
on the half line. In spite of this fact, the four-point function 
(5.17) differs from the free one. We would like to recall 
in this respect that according to Jost's theorem (see e.g. Ref. 16), 
such a phenomenon is forbidden in relativistic invariant models.

\newchapt {VI. SCATTERING THEORY}

As it is well known, integrable quantum systems on the real line 
are characterized by a factorized scattering matrix. This means 
that multiparticle scattering is described by an appropriate product 
of two-particle scattering matrices, which in turn are subject to 
physical constraints like unitarity, crossing symmetry, etc.  

Some years ago, Cherednik$^{12}$ proposed a version of factorized 
scattering, adapted to the half line case. The following physical 
picture emerges from his investigation. 
Let $|k_1,...,k_n \rangle^{\rm in}$ be an in-state,
representing $n$ particles coming from $x=+\infty$ and
thus having negative momenta $k_1<k_2<...<k_n<0$.
These particles interact among themselves before and after being
reflected by the wall at $x=0$, giving rise to
an out-state $|p_1,...,p_m \rangle^{\rm out}$
composed of particles traveling towards $x=+\infty$ and thus having
positive momenta $p_1>p_2>...>p_m>0$.
The transition amplitude between these states vanishes unless
$n=m$ and $p_i = -k_i$, $i=1,...,n$. Therefore, not only
the total momentum, but each momentum is separately reflected.
According to Ref. 12, the scattering amplitude is
$$
{}^{\rm out} \langle p_1,...,p_m |k_1,...,k_n \rangle ^{\rm in} =
\delta_{mn} \prod_{i=1}^n 2\pi \delta(p_i +k_{i})B(p_i)
\prod_{i,j=1 \atop i<j}^n R(p_i,p_j)R(p_i,-p_j) \; . \neweq 
$$ 
The $R$-factors describe the interactions among the
particles in the bulk, while the\break $B$-factors take into 
account the reflection from the wall. 

The main goal of this section is to prove that the NLS model 
on $\hl $ perfectly fits the scheme of Cherednik. In order to 
do that, we must develop first the scattering theory 
corresponding to the off-shell quantum field 
$\Phi^*(t,f)$. Our framework will be the conventional Haag-Ruelle 
approach$^{17}$, suitably adapted to the nonrelativistic case. 

A first relation between the quantum solution (4.6,7) and 
Cherednik's scattering amplitude (6.1) is obtained 
through the identification
$$
|p_1,...,p_n\rangle^{\rm out} = a^*(p_1)...a^*(p_n)\Omega \; ,
\quad  p_1>...>p_n>0 \; , \neweq 
$$
$$ 
|k_1,...,k_n\rangle^{\rm in} = a^*(k_1)...a^*(k_n)\Omega \; ,
\quad  k_1<...<k_n<0 \; . \neweq  
$$ 
We recall in fact that $\balg $ has been designed in such a way, 
that the amplitudes 
$$
\langle a^*(p_1)...a^*(p_m)\Omega \, ,\, 
a^*(k_1)...a^*(k_n)\Omega \rangle \; , \neweq 
$$ 
precisely reproduce the right hand side of Eq.(6.1). What is 
still missing therefore is the construction 
of suitable states, expressed in terms of 
$\Phi^*(t,h)$ and $\Omega $, which approach the out-states 
(6.2) for $t \to \infty $ and the in-states (6.3) for 
$t \to -\infty $. We are now going to fill this gap. 

Proposition 5 shows that $\Phi^* (t,f)$, restricted 
on $\H_{R,B}^n$ is a bounded operator of norm 
$$
\Vert \Phi^*(t,f) \Vert  \leq \sqrt{n+1}\, \Vert f \Vert \, ,
\neweq
$$
which in turn implies that it can be extended to any $f\in L^2(\hl )$. 
{}From the estimates (3.21) we know that also $a^*(h) $ is 
bounded on $\H^n_{R,B}$, where 
$$
\Vert a^*(h) \Vert \leq \sqrt {n+1}\, \Vert h \Vert \, , 
\qquad \forall \, h\in L^2 (\R ) 
\, . \neweq 
$$
Combining this inequality with the definition (4.6), one finds, 
$$
\Vert \int_{\R^n} dx_1...dx_n\, f(x_1,...,x_n)
\at^*(t,x_1)\cdots \at^*(t,x_n) \Omega \Vert \leq 
\sqrt {n!}\, \Vert f\Vert \, , \qquad \forall\, f\in L^2 (\R^n ) \, . 
\neweq 
$$

In order to develop the Haag-Ruelle formalism, we will need also 
the following notations. Let $h(k) \in \S(\R )$. Then we set:  
$$
h^t (x) \equiv \int_\R {dk \over 2\pi}\e^{ i k x - i k^2 t} h(k) 
\, , \qquad h^t_+(x) \equiv \theta (x) [h^t(x) + h^t(-x)] \, , 
\qquad \widetilde h (k) \equiv h(-k) \, , \neweq 
$$
where $\theta (x)$ is the Heaviside step function. Notice that 
$$
\ht^t_+(x) = \theta (x)[\ht^t(x) + \ht^t(-x)] = 
\theta (x) [h^t(-x) + h^t(x)] = h^t_+(x) \, . \neweq 
$$
We are now in position to formulate 
\medskip 
\noindent {\bf Theorem 7:} (Asymptotic states) {\it Let 
$$
h_1 \succ h_2 \succ \cdots \succ h_n \, , \qquad h_j \in \S(\hl ) \, , 
\quad j =1,...,n \, . 
$$
Then one has the following strong limits}
$$
\lim_{ t\to + \infty }
\Phi^*(t, h^t_{1+})\Phi^* (t, h^t_{2+}) \cdots \Phi^*(t, h^t_{n+})\Omega 
= a^*(h_1) a^*(h_2) \cdots a^*(h_n) \Omega \, , \neweq 
$$
$$ 
\lim_{ t\to - \infty }
\Phi^*(t, h^t_{1+}) \Phi^* (t, h^t_{2+}) \cdots \Phi^*(t, h^t_{n+}) \Omega 
= a^*(\ht_1) a^*(\ht_2) \cdots a^*(\ht_n) \Omega \, . \neweq 
$$ 
\medskip 

\noindent For proving this statement, we need some preliminary 
results.

\medskip
\noindent {\it Lemma 4: Let $h\in \S(\hl )$. Then}
$$
\lim_{t\to +\infty} \Vert h^t_+ - h^t \Vert = 0
\, , \qquad 
\lim_{t\to -\infty} \Vert h^t_+ - \ht^t \Vert = 0 \, . 
\neweq 
$$
\medskip 
\noindent {\it Proof:} A direct computation gives: 
$$
\Vert h^t_+ - h^t \Vert^2 = 
2i\int_{\R^2} {dk\over 2\pi} {dp\over 2\pi}\, \overline h(k) h(p)  
\, {\e^{it(k+p)(k-p)}\over k-p+i\epsilon } \, , \neweq 
$$
$$
\Vert h^t_+ - \ht^t \Vert^2 =  
-2i\int_{\R^2} {dk\over 2\pi} {dp\over 2\pi}\, \overline h(k) h(p)
\, {\e^{it(k+p)(k-p)}\over k-p-i\epsilon } \, . \neweq 
$$
{}Now, for proving Eq.(6.12), it is enough to take into account 
that $\supp h > 0$ and to use the weak limit 
$$
\lim_{t\to \pm \infty}{\e^{itk}\over k \pm i\epsilon } = 0 \, . 
\neweq 
$$ 
\medskip 

\noindent {\it Corollary 1: Let $h_1,h_2,\ldots,h_n\in \S(\bfR_+)$. Then} 
$$
\lim_{t\to +\infty } \Vert h^t_{1+} \otimes \cdots \otimes h^t_{n+} -
h^t_1 \otimes \cdots \otimes h^t_n  \Vert = 0
\, , \neweq
$$
$$
\lim_{t\to -\infty } \Vert h^t_{1+} \otimes \cdots \otimes h^t_{n+} -
\ht^t_1 \otimes \cdots \otimes \ht^t_n \Vert = 0
\, . \neweq
$$
\medskip 

\noindent {\it Lemma 5: Let $h_1, h_2 \in \S(\hl )$ are such that
$h_1 \succ h_2$. Then, the functions 
$$
H^t(x_1,x_2) = h^t_1(x_1) h^t_2(x_2) \, \theta (x_2-x_1) 
\, , \neweq
$$
$$
\widetilde H^t(x_1,x_2) = \ht^t_1(x_1) \ht^t_2(x_2)\, \theta (x_2-x_1) 
\, , \neweq
$$
satisfy}
$$
\lim_{t\to +\infty} \Vert H^t \Vert = 0 \, , \qquad  
\lim_{t\to -\infty} \Vert \widetilde H^t \Vert = 0 \, .
\neweq
$$
\medskip 

\noindent {\it Proof:} Let us consider for instance $H^t$. One has 
$$
\Vert H^t \Vert^2 = 
\int_{-\infty}^{\infty}  dx_1 \int^\infty_{x_1} dx_2
\vert h^t_1 (x_1) h^t_2 (x_2) \vert ^2 = 
$$
$$
\int_{\R^4} {dk_1\over 2\pi}{dp_1\over 2\pi}{dk_2\over 2\pi}{dp_2\over 2\pi} 
\, {\overline h}_1(k_1) h_1(p_1) {\overline h}_2(k_2) h_2(p_2) 
I(k_1,p_1,k_2,p_2) \e^{(k_1^2 -p_1^2 +k_2^2 -p_2^2)t} \, , \neweq 
$$
with  
$$
I(k_1,p_1,k_2,p_2) \equiv 
\int_{-\infty}^{\infty}  dx_1 \int^\infty_{x_1} dx_2 \,
\e^{i\left [(p_1 - k_1)x_1 + (p_2-k_2) x_2 \right ]} \, . 
$$
The integration in $x_1$ and $x_2$ gives 
$$
I(k_1,p_1,k_2,p_2) = 
{2\pi i \delta(k_1 - p_1 + k_2-p_2) \over p_2 - k_2 + i \epsilon} 
\, . \neweq 
$$
Therefore, 
$$
\Vert H^t \Vert^2 = 
i\int_{\R^3} {dp_1\over 2\pi}{dp_2\over 2\pi}{dk_2\over 2\pi}
\, {\overline h}_1(p_1-p_2+k_2) h_1(p_1) {\overline h}_2(k_2) h_2(p_2) 
{\e^{2i(p_1-k_2)(p_2-k_2)t} \over p_2 - k_2 + i\epsilon }\, .  
\neweq 
$$
The support properties of the function $h_1$ and $h_2$ imply that 
the integrand vanishes unless $p_1 > k_2 >0$, which completes 
the argument because of Eq.(6.15). Analogous considerations apply 
to $\widetilde H^t$. 

\medskip 

\noindent {\it Corollary 2: Let 
$$
G^t (x_1,x_2) = h^t_{+1}(x_1) h^t_{+2}(x_2)\, \theta (x_2-x_1) 
\, . \neweq
$$
Then}
$$
\lim_{t\to \pm \infty} \Vert G^t \Vert = 0 \, . \neweq 
$$
\medskip 
  
\noindent {\it Proof:} One has to combine Eqs.(6.9,16,17,20). 

\medskip 

The statement of Corollary 2 has the following 
generalization to the case of\break $n\geq 2$ variables. Suppose that 
$h_1,...,h_n \in \S(\hl )$ and $h_1\succ ... \succ h_n$. 
Let $\P_n $ be the group of all permutations of the indices 
$\{1,2,...,n \}$. For any $\sigma \in \P_n$ we define the function  
$$
G^t_\sigma (x_1,...,x_n) \equiv 
 h^t_{1+}(x_1)\cdots h^t_{n+}(x_n) \theta (x_{\sigma_1},..., x_{\sigma_n}) 
\, , \neweq 
$$
where 
$$
\theta (x_{\sigma_1},...,x_{\sigma_n}) \equiv \prod_{i,j = 1\atop i<j}^n 
\theta (x_{\sigma_i}-x_{\sigma_j}) \, . \neweq 
$$
\medskip 

\noindent {\it Corollary  3: For any $\sigma \in \P_n$ different 
from the identity $e=(1,2,...n)$, one has} 
$$
\lim_{t\to \pm \infty }\Vert G^t_\sigma \Vert = 0 
\, . \neweq 
$$
\medskip 

We are now ready to prove Theorem 7. 
\medskip 

\noindent {\it Proof:} The case $n=1$ is quite simple. 
Using the identities  
$$
\Phi^*(t,f) \Omega = \at^*(t,f)\Omega \, , \qquad 
a^*(h) = \at^*(t,h^t) \quad ,\neweq 
$$
one finds 
$$
\Vert \Phi^* (t, h^t_+)\Omega - a^*(h)\Omega \Vert =
\Vert \at^*(t, h^t_+)\Omega - \at^*(t, h^t)\Omega \Vert 
\leq \Vert h^t_+ - h^t \Vert  \, , \neweq  
$$
which according to Lemma 5 tends to $0$ in the limit 
$t\to +\infty $. Let us consider now the case $n\geq 2$. 
Applying Eq.(4.34) and 
$$
\theta (x_1,...,x_n) = 1 - 
\sum_{\sigma \in \P_n \atop \sigma \not= e} 
\theta (x_{\sigma_1},..., x_{\sigma_n}) \, , \neweq 
$$
we get: 
$$
\Phi^*(t, h^t_{1+}) \cdots \Phi^*(t, h^t_{n+})\Omega = 
$$
$$
\int_{\R^n} dx_1...dx_n h^t_{1+}(x_1) \cdots h^t_{n+}(x_n) 
\sum_{\sigma \in \P_n}\theta (x_{\sigma_1},..., x_{\sigma_n}) 
\at^*(t,x_{\sigma_1})\cdots \at^*(t,x_{\sigma_n}) \Omega =
$$
$$
\at^*(t,h^t_{1+}) \cdots \at^*(t,h^t_{n+})\Omega + 
$$
$$
\sum_{\sigma \in \P_n \atop \sigma \not= e} 
\int_{\R^n} dx_1...dx_n G^t_\sigma (x_1,...,x_n) 
\left [\at^*(t,x_{\sigma_1})\cdots \at^*(t,x_{\sigma_n}) \Omega 
- \at^*(t,x_1)\cdots \at^*(t,x_n) \Omega \right ] 
\, . \neweq 
$$
The estimate (6.7) then leads to  
$$
\Vert \Phi^*(t, h^t_{1+}) \cdots \Phi^*(t, h^t_{n+})\Omega - 
a^*(h_1)\cdots a^*(h_n)\Omega \Vert = 
$$
$$
\Vert \Phi^*(t, h^t_{1+}) \cdots \Phi^*(t, h^t_{n+})\Omega - 
\at^*(t,h^t_1)\cdots \at^*(t,h^t_n)\Omega \Vert \leq 
$$
$$
\Vert \at^*(t,h^t_{1+}) \cdots \at^*(t,h^t_{n+})\Omega 
- \at^*(t,h^t_1)\cdots \at^*(t,h^t_n)\Omega \Vert + 
2\sqrt {n!}\, \sum_{\sigma \in \P_n \atop \sigma \not= e} 
\Vert G_\sigma^t \Vert \leq 
$$
$$
\sqrt {n!} \Vert h^t_{1+} \otimes \cdots \otimes h^t_{n+} -
h^t_1 \otimes \cdots \otimes h^t_n  \Vert  + 
2\sqrt {n!} \sum_{\sigma \in \P_n \atop \sigma \not= e} 
\Vert G_\sigma^t \Vert  \, , \neweq 
$$
which implies the strong limit (6.10). Analogous considerations 
give 
$$
\Vert \Phi^*(t, h^t_{1+}) \cdots \Phi^*(t, h^t_{n+})\Omega - 
a^*(\ht_1)\cdots a^*(\ht_n)\Omega \Vert \leq  
$$
$$
\sqrt {n!} \Vert h^t_{1+} \otimes \cdots \otimes h^t_{n+} -
\ht^t_1 \otimes \cdots \otimes \ht^t_n  \Vert  + 
2\sqrt {n!}\, \sum_{\sigma \in \P_n \atop \sigma \not= e} 
\Vert G_\sigma^t \Vert  \, , \neweq 
$$
which proves (6.11). 
\medskip 

We proceed with the construction of the scattering operator 
$S$, following the general strategy developed in Ref. 18. 
According to Theorem 7, the asymptotic spaces $\F^{{\rm out}}$ and 
$\F^{{\rm in}}$ are generated by finite linear combinations 
of the vectors ($n\geq 1$)  
$$
\E^{{\rm out}} = 
\{\, \Omega ,\,  a^\ast (h_1) \cdots a^\ast (h_n) \, \Omega \, \, : 
\, \, h_1\succ \cdots \succ h_n ,\, \, h_j \in \S(\hl )\, \} \neweq 
$$
and 
$$
\E^{{\rm in}} = 
\{\, \Omega ,\, a^\ast (\widetilde h_1) \cdots 
a^\ast (\widetilde h_n) \, \Omega \, \, : \, \, 
h_1\succ \cdots \succ h_n, \, \, h_j\in \S(\hl )\, \}  
\neweq 
$$
respectively. One can show$^{11}$ moreover, that $\F^{{\rm out}}$ 
and $\F^{{\rm in}}$ are separately dense $\brep $. 
This property of asymptotic completeness allows to 
demonstrate$^{11}$ that the mapping 
$S\, :\, \E^{{\rm out}} \rightarrow \E^{{\rm in}}$, defined by 
$$
S\Omega = \Omega \quad , \neweq 
$$
$$
S\, a^\ast (h_1) a^\ast (h_2)\cdots a^\ast (h_n)\Omega
= a^\ast (\widetilde h_1) a^\ast (\widetilde h_2)\cdots 
a^\ast (\widetilde h_n)\Omega 
\quad , \neweq 
$$
extends to a unitary scattering operator on $\brep $. 
We stress that $S$ is nontrivial, in spite of 
the fact that the quantum fields $\Phi $ and $\Phi^*$ 
realize a Fock representation of the canonical commutation 
relations. This feature is not in contradiction with Haag's 
theorem$^{16}$, because we are dealing with a nonrelativistic 
system, which does not satisfy in particular relativistic 
local commutativity.  

The construction of the scattering operator $S$ 
completes the picture and concludes our quantum field 
theory description of the NLS model on ${\bf R}_+$.

\newchapt {VII. OUTLOOK AND CONCLUSIONS}

We studied the nonlinear Schr\"odinger equation on the half 
line with mixed boundary condition. After a brief discussion 
of some aspects of the corresponding classical boundary value problem, 
we constructed the exact second quantized solution of the system, 
establishing its basic properties. The explicit form of our solution 
shows that the quantum inverse scattering transform works 
also on the half line, provided that the Zamolodchikov-Faddeev algebra is 
replaced by the boundary algebra ${\cal B}_R$. This is one of the main 
results of the present paper. It demonstrates that besides being an 
useful tool in scattering theory$^{11}$, the concept of boundary algebra 
is essential also for the construction of off-shell interacting fields in 
integrable systems on $\hl $. We emphasize in this respect, 
that our results have a straightforward generalization 
to all elements of the NLS hierarchy (e.g. the complex modified 
Korteveg-de Vries equation) on the half line. The case with internal 
$SU(N)$ symmetry can also be treated analogously. 

As for future extensions of the present work, it would be 
interesting to investigate the range $\eta < 0$. The new phenomenon, 
which can be expected on general grounds, is 
the presence of boundary bound states. Taking into account 
that one can describe by $\balg $ also degrees of freedom residing 
on the boundary (see the appendix of Ref. 11), we strongly believe that 
our framework extends to the case $\eta < 0$ as well.

\vfill \eject 

\centerline {\bbf REFERENCES} 

\bigskip 
\medskip 

\item {$^1$} L. D. Faddeev and L. A. Takhtajan, {\it Hamiltonian 
Methods in the Theory of Solitons} (Springer-Verlag, Berlin, 1987).  

\item {$^2$} E. Sklyanin, J. Phys. A: Math. Gen. {\bf 21}, 2375 (1988). 

\item {$^3$} A. S. Fokas, Physica D {\bf 35}, 167 (1989). 

\item{$^4$} E. Sklyanin and L. D. Faddeev, Sov. Phys. Dokl.
{\bf 23}, 902 (1978); E. Sklyanin, {\it ibid.} {\bf 24}, 107 (1979).

\item{$^5$} H. B. Thacker and D. Wilkinson, Phys. Rev. D {\bf 19}, 
3660 (1979); D. B. Creamer, H. B. Thacker and D. Wilkinson,
{\it ibid.} {\bf 21}, 1523 (1980).

\item{$^6$} H. B. Thacker, in {\it Integrable Quantum Field Theories}, 
edited by J. Hietarinta and C. Montonen (Springer-Verlag, Berlin, 1982).

\item{$^7$} J. Honerkamp, P. Weber and A. Wiesler, Nucl. Phys. B
{\bf 152}, 266 (1979).

\item{$^8$} B. Davies, Journ. Phys. A: Math. Gen. {\bf 14}, 2631 (1981);
Inverse Problems {\bf 4}, 47 (1988). 

\item{$^9$} E. Gutkin, Phys. Rep. {\bf 167}, 1 (1988). 
 
\item{$^{10}$} A. B. Zamolodchikov and A. B. Zamolodchikov, Ann. Phys.
{\bf 120}, 253 (1979); L. D. Faddeev, Sov. Scient. Rev. Sect. C
{\bf 1}, 107 (1980). 

\item{$^{11}$} A. Liguori, M. Mintchev and L. Zhao, 
Commun. Math. Phys. {\bf 194}, 569 (1998). 

\item{$^{12}$} I. V. Cherednik, Theor. Math. Phys. {\bf 61}, 977 (1984). 

\item{$^{13}$} M. Gattobigio, A. Liguori and M. Mintchev, Phys. Lett. B 
{\bf 428}, 143 (1998). 

\item{$^{14}$} R. R. Rosales, Stud. Appl. Math. {\bf 59}, 117 (1978). 

\item{$^{15}$} A. Liguori and M. Mintchev, Commun. Math. Phys. {\bf 169}, 
635 (1995); Lett. Math. Phys. {\bf 33}, 283 (1995).

\item{$^{16}$} R. F. Streater and A. S. Wightman, {\it PCT, Spin and Statistics, 
and All That} (Addison-Wesley, 1980). 

\item{$^{17}$} M. Reed and B. Simon, {\it Methods of Modern Mathematical 
Physics III: Scattering Theory} (Accademic Press, New York 1979). 

\item{$^{18}$} A. Liguori, M. Mintchev and M. Rossi, J. Math. Phys. 
{\bf 38}, 2888 (1997).

\vfill\eject 
\bye